\documentclass[twocolumn,twoside,slac]{revtex4}
\usepackage{graphicx}
\usepackage{fancyhdr}
\begin{document}

\title{The DZERO Level 3 Data Acquistion System}

%

\author{D. Chapin, M. Clements, D. Cutts, S. Mattingly}
\affiliation{Department of Physics, Brown University, Providence, RI 02912 USA}
\author{B. Angstadt, G. Brooijmans, D. Charak, S. Fuess, A. Kulyavtsev, M. Mulders, D. Petravick, R. Rechenmacher, D. Zhang}
\affiliation{FNAL, Batavia, IL 60510, USA}
\author{R. Hauser}
\affiliation{Department of Physics, Michigan State University, East Lansing, MI 48824 USA}
\author{P. Padley}
\affiliation{TW Bonner Nuclear Lab, Rice University, Houston, TX 77251 USA}
\author{A. Haas, D. Leichtman, G. Watts}
\affiliation{Department of Physics, University of Washington, Seattle, WA 98195, USA}

\begin{abstract}
The DZERO experiment located at Fermilab has recently started RunII with an upgraded detector. The RunII physics program requires the Data Acquisition to readout the detector at a rate of 1 KHz. Events fragments, totaling 250 KB, are readout from approximately 60 front end crates and sent to a particular farm node for Level 3 Trigger processing. 
A scalable system, capable of complex event routing, has been designed 
and implemented based on commodity components: 
VMIC 7750 Single Board Computers for readout, a Cisco 6509 switch for 
data flow, and close to 100 Linux-based PCs for high-level event filtering.


\end{abstract}

\maketitle

\thispagestyle{fancy}


\section{Introduction}
The Level 3 Data Acquistion System 
(L3DAQ)~\cite{L3DAQ_RT2003_PROCEEDINGS,L3DAQ_RT2003_JOURNAL} 
transports detector component data located in VME readout crates 
to the processing nodes of the Level 3 trigger filtering farm.
At a rate of 1kHz, sixty-three VME crates must be
read out for each event, each containing 1-20~kB of data distributed among VME
modules. The total event size is approximately 250 kilobytes. 

\begin{figure*}
\centering
\includegraphics[height=6.0in,angle=-90]{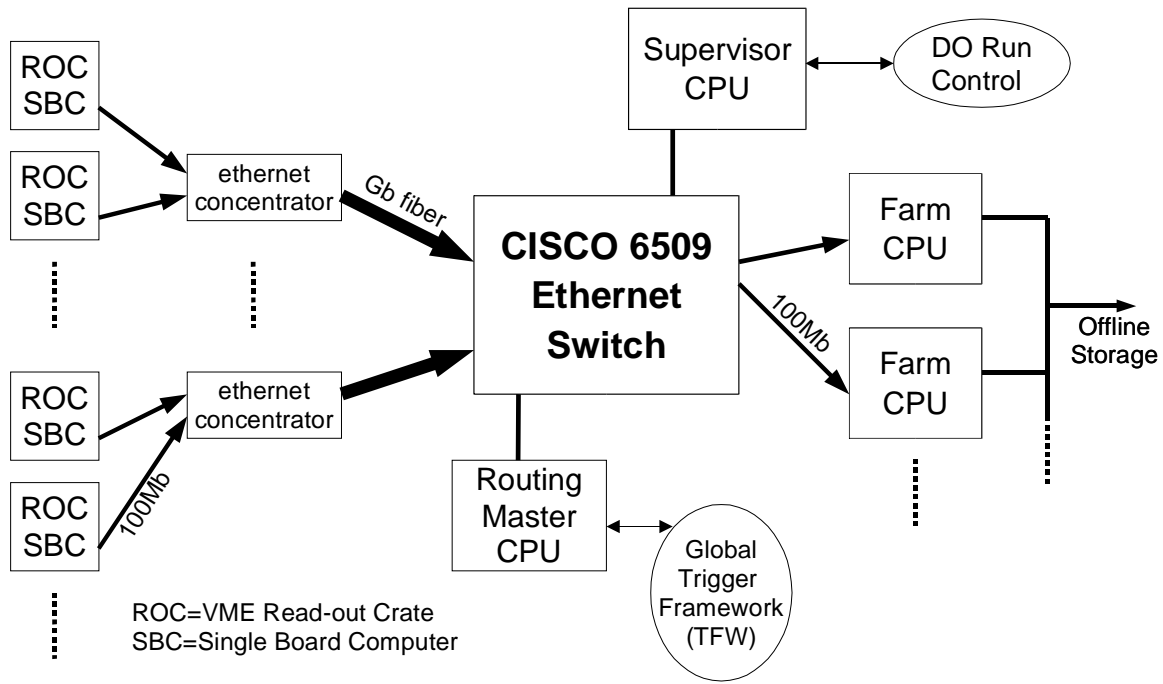}
\caption{The physical network configuration of the L3DAQ system.}
\label{fig:hardware}
\end{figure*}
\begin{figure*}
\centering
\includegraphics[height=5.6in,angle=-90]{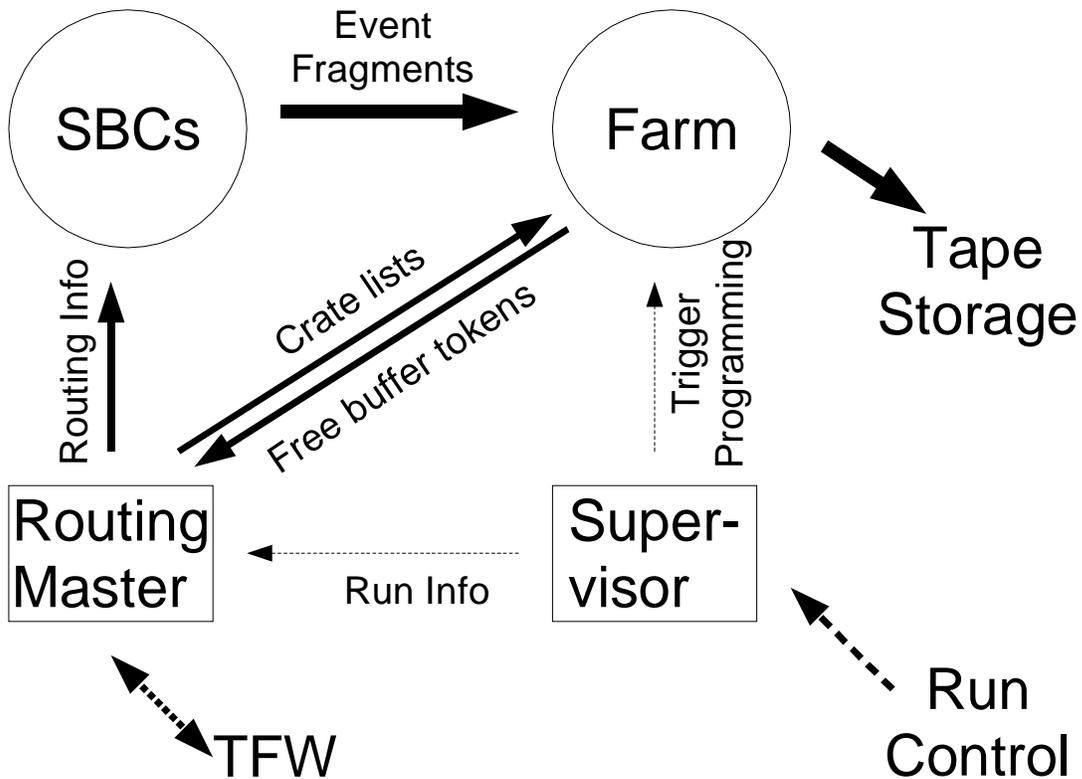}
\caption{Schematic illustration of the information and 
dataflow through the L3DAQ system.}
\label{fig:dataflow}
\end{figure*}

As shown in figure \ref{fig:hardware}, the system is built around 
a single Cisco 6509\cite{CISCO} ethernet switch.
A schematic of the communication and data flow in the system is shown
in figure~\ref{fig:dataflow}. All nodes in the system are based on 
commodity computers and run the Linux operating system.
TCP/IP sockets implemented via the ACE~\cite{ACE} C++ network and
utility library
are used for all communication and data transfers.

\section{Operation}
The Supervisor process provides the interface between the 
main D0 run control (COOR) and the L3DAQ system. 
When a new 
run is configured, the Supervisor passes run and general
trigger information to the RM and passes the COOR-provided L3 filter 
configuration to the IO/EVB process on relevant farm nodes, 
where it is cached and passed on to the Level 3 filter processes.

A single-board computer (SBC) in each VME crate reads out the VME
modules and sends the data to one or more farm nodes
specified by routing instructions received from the Routing Master (RM)
process. An Event Builder (EVB) process on each farm node builds a complete
event from the received event fragments and makes it available to 
Level 3 trigger filter processes.

The SBCs are 
single-board computers with
dual 100~Mb/s Ethernet interfaces and a 
VME-to-PCI interface.
An expansion slot is occupied by a 
digital-I/O (DIO) module, 
used to coordinate the readout of VME modules over the VME user (J3) backplane.
A custom kernel driver on the SBC handles
interrupt requests from the DIO module which are triggered by
readout requests from the crate-specific electronics.
On each readout request the kernel module 
performs the VME data transfers and stores the event
fragment in one 
of several buffers in kernel-memory.

A user-level process on the SBC receives  
route information from the Routing Master 
in the form of Route Tags that contain a
unique event identifier (L3 transfer number) and the indices of
the farm nodes to
which that event should be sent. 
If the Route Tag's L3 transfer number matches that of the transfer number 
embedded within the head event fragment in the kernel buffers, the event
fragment is sent to the specified farm nodes. 

An Event Builder process (EVB) on each farm node
collates the event fragments received from SBCs
into complete events, keyed by L3 transfer number. 
For each event the EVB receives an
expected-crate list from the RM in order to determine when
an event is complete. 
Complete events
are placed in shared memory event buffers 
for processing by the Level 3 filter processes.
The EVB routinely informs the RM of the number of
available event buffers that it has, so that the RM
will not route an event to a farmnode unless the
event can be processed immediately.

The Routing Master program executes on an SBC in a special VME crate
which contains a hardware interface to the D0 trigger framework (TFW).
The TFW provides trigger information and the L3 transfer number for
each event and allows the RM to asynchronously disable the firing
of triggers.
For each event the RM program chooses a node for processing based on 
the run configuration, 
the trigger information, and the number of available buffers 
in the set of nodes configured to process the type of event.  
A node is chosen in a round-robin fashion from amongst the set of nodes
with the most free buffers.
If the number of available free buffers is too low, the RM instructs the
TFW to disable triggers so that the farm nodes have time to catch up.

To avoid dropped packets in the main switch, data flow 
is limited by setting the TCP/IP receive window size 
on the farmnodes. The window size is set such that
the product of the number of connection sources and
the receive window size is smaller than the switch's 
per-port output buffer memory.

\section{Conclusion}
The DZERO Level 3 data acquistion system has been built from 
commercially available hardware: single board VME computers, ethernet
switches, and PCs. The software components rely upon
high level programming languages; the Linux operating system;
widely-used, open libraries; and standard networking protocols.
The system has performed reliably since commissioning in May 2002.
Additional details are availble from the 
references~\cite{L3DAQ_RT2003_JOURNAL}.

\end{document}